\documentclass[10pt,english]{article}
\usepackage{times}
\usepackage[T1]{fontenc}
\usepackage[latin1]{inputenc}
\usepackage{amsmath}
\usepackage{amssymb}

\makeatletter

\providecommand{\tabularnewline}{\\}

\usepackage{slashed}

\usepackage{babel}
\makeatother
\begin{document}

\title{\textbf{Neutral currents in a $SU(4)_{L}\otimes U(1)_{Y}$ gauge
model with exotic electric charges }}

\author{ADRIAN PALCU}

\date{\emph{Faculty of Exact Sciences - {}``Aurel Vlaicu'' University
Arad, Str. Elena Dr\u{a}goi 2, Arad - 310330, Romania}}

\maketitle
\begin{abstract}
The weak currents with respect to the diagonal neutral bosons $Z$,$Z^{\prime}$,
and $Z^{\prime\prime}$ of a specific \textbf{$SU(4)_{L}\otimes U(1)_{Y}$}
gauge model are computed in detail for all the fermion families involved
therein. Our algebraical approach, which is based on the general method
of solving gauge models with high symmetries proposed several years
ago by Cot\u{a}escu, recovers in a non-trivial way all the Standard
Model values for current couplings of the traditional leptons and
quarks, and predicts plausible values for those of the exotic fermions
in the model. 


\end{abstract}

\section{Introduction}

In this letter we focus on the neutral currents of a specific gauge
model based on the enlarged local group \textbf{$SU(3)_{C}\otimes SU(4)_{L}\otimes U(1)_{Y}$}
(in short, 3-4-1). In the last decade, this has been a very visited
extension of the Standard Model \cite{key-1} - \cite{key-3} for
it can accomodate in a natural manner some new phenomena such as neutrino
oscillation (with resulting tiny masses) or extra neutral bosons to
be detected at LHC. As usual, the electro-weak sector undergoes a
spontaneous symmetry breakdown (SSB) up to the $U(1)_{em}$ residual
one of electromagnetic interaction. Although, some older versions
of the left-right symmetric models based on the gauge group $SU(2)_{L}\otimes SU(2)_{R}\otimes U(1)_{B-L}$
\cite{key-4} - \cite{key-7} seem to presume the $SU(4)_{L}\otimes U(1)_{Y}$
electro-weak approach, the latter has been in its own right largely
studied \cite{key-8} - \cite{key-22} by taking into account both
its {}``exotic electric charge'' version and {}``no exotic electric
charge'' version. This split can be made on different ways \cite{key-18}
- \cite{key-20}, but here we resort to the algebraical method proposed
by Cot\u{a}escu \cite{key-23} and worked out by the author in some
recent papers \cite{key-17,key-20,key-24} on 3-4-1 models.

\paragraph{General method}

The general method mainly consits of a particular geometrization of
the scalar sector of the gauge model of interest. Since, the local
gauge group $SU(N)_{L}\otimes U(1)_{Y}$ is taken into consideration,
the fermion families will be nothing but the irreducible unitary representations
(irrep) of this group (with a proper hypercharge assignment so that
all the anomalies cancel) and the gauge bosons (associated to the
generators of the corresponding Lie algebra of the symmetry group)
couple the fermions through covariant derivatives in the usual manner.
Also, a minimal Higgs mechanism (mHm) is designed to spontaneously
break this symmetry up the the universal electromagnetic one ($U(1)_{em}$).
The above mentioned geometrization of the scalar sector is intended
to supply - once the SSB took place - the non-degenerate boson mass
spectrum. A step further, one can obtain the general expressions for
the electric and neutral currents \emph{- i.e}. the electric and neutral
charges of the fermions with respect to all physical bosons in the
model - by just setting a versor choice in a general Weinberg transformation
(gWt) that makes the job of reaching the physical basis from the gauge
one through a special rotation $\omega\in SO(N-1),$as the electromagnetic
field $A_{\mu}^{em}$has been separated. That is, $\omega$ diaonalizes
the mass matrix of the neutral bosons and it takes particular shapes
from one particular model to another.

All the details of the general approach can be found in Ref. \cite{key-23}
and they are taken for granted without insisting here on deducing
them one more time. The resulting formuale are finally adapted to
the particular model we work on.

\paragraph{Electric and Neutral Charges}

The charges of the particles are idetified with the resulting coupling
coefficients of the currents with respect to the well-determined physical
bosons. We are interested in diagonal bosons (those associated with
diagonal generators of the gauge group - $D^{\rho}$- in the irrep
$\rho$, in our notation). Hence, according to Eqs. (58) - (59) in
Ref. \cite{key-23}, the spinor multiplet $L^{\rho}$ (of the irrep
$\rho$) presents the following electric charge matrix \begin{equation}
Q^{\rho}=g\left[(D^{\rho}\cdot\nu)\sin\theta+y_{\rho}\cos\theta\right],\label{Eq.1}\end{equation}
 and the $n-1$ neutral charge matrices \begin{equation}
Q^{\rho}(Z^{i})=g\left[D_{k}^{\rho}-\nu_{k}(D^{\rho}\cdot\nu)(1-\cos\theta)-y_{\rho}\nu_{k}\sin\theta\right]\omega_{\cdot\; i}^{k\;\cdot},\label{Eq.2}\end{equation}
 corresponding to the $n-1$ neutral physical fields, $Z_{\mu}^{i}$.
We note that all the other gauge fields, namely the charged bosons
$A_{j\mu}^{i}$, have the same coupling, $g/\sqrt{2}$, to the fermion
multiplets.

\section{$SU(4)_{L}\otimes U(1)_{Y}$ model with exotic electric charges}

We work here within the framework of the particular 3-4-1 model with
exotic electric charges (see Ref. \cite{key-20,key-24} for the electric
charge assignment), based on the versor setting $\nu_{1}=1$, $\nu_{2}=0$,
$\nu_{3}=0$ - Case 1 in the Ref. \cite{key-20}. In what follows
we denote the irreps by $\rho=(\mathbf{n}_{color},\mathbf{n}_{\rho},y_{ch}^{\rho})$
indicating the genuine chiral hypercharge $y_{ch}$ instead of $y$
of the general method. This replacement $y\rightarrow\frac{g^{\prime}}{g}y_{ch}$
exploits a specific relation among the two gauge couplings $g$ (of
the $SU(N)_{L}$) and $g^{\prime}$ (of the $U(1)_{Y}$). Evidently,
it was initially designed to simplify the work by dealing with only
one gauge coupling (instead of two) in computations of the general
method.

\paragraph{Fermion sector}

With these assumptions, the multiplets of the 3-4-1 model of interest
here will be denoted by $(\mathbf{n}_{color},\mathbf{n}_{\rho},y_{ch}^{\rho})$,
and one gets the following particle content. 

\textbf{Lepton families}\begin{equation}
\begin{array}{ccccc}
f_{\alpha L}=\left(\begin{array}{c}
e_{\alpha}^{c}\\
e_{\alpha}\\
\nu_{\alpha}\\
N_{\alpha}\end{array}\right)_{L}\sim(\mathbf{1,4},0)\end{array}\label{Eq.3}\end{equation}

\textbf{Quark families}\begin{equation}
\begin{array}{ccc}
Q_{iL}=\left(\begin{array}{c}
J_{i}\\
u_{i}\\
d_{i}\\
D_{i}\end{array}\right)_{L}\sim(\mathbf{3,4^{*}},-1/3) &  & Q_{3L}=\left(\begin{array}{c}
J_{3}\\
-b\\
t\\
T\end{array}\right)_{L}\sim(\mathbf{3},\mathbf{4},+2/3)\end{array}\label{Eq.4}\end{equation}
\begin{equation}
\begin{array}{ccc}
(b_{L})^{c},(d_{iL})^{c},(D_{iL})^{c}\sim(\mathbf{3},\mathbf{1},-1/3) &  & (t_{L})^{c},(u_{iL})^{c},(T_{L})^{c}\sim(\mathbf{3},\mathbf{1},+2/3)\end{array}\label{Eq.5}\end{equation}
\begin{equation}
\begin{array}{ccccccccc}
(J_{3L})^{c}\sim(\mathbf{3,1},+5/3) &  &  &  &  &  &  &  & (J_{iL})^{c}\sim(\mathbf{3,1},-4/3)\end{array}\label{Eq.6}\end{equation}
 with $\alpha=1,2,3$ and $i=1,2$. Capital letters denote exotic
fermions, while small ones the fermions predicted by the SM. One can
easily check that all the anomalies cancel by an interplay among families.

In addition, the connection between the $\theta$ angle of our method
\cite{key-23} and $\theta_{W}$(the Weinberg angle from SM) was inferred
\cite{key-20}: $\sin\theta=2\sin\theta_{W}$ that points to the coupling
relation: $g^{\prime}/g=\sin\theta_{W}/\sqrt{1-4\sin^{2}\theta_{W}}$. 

As in all cases when the general method was applied to particular
3-4-1 models \cite{key-17,key-20,key-24} the first gauge coupling
$g$ is identical to the SM first coupling.

\paragraph{Boson sector}

The electro-weak boson sector is essentially determined by the generators
of the Lie algebra associated to the gauge group $SU(4)_{L}\otimes U(1)_{Y}$.
We use here the standard generators $T_{a}$ of the $su(4)$ algebra,
so that the Hermitian diagonal generators of the Cartan subalgebra
are, in order, $D_{1}=T_{3}=\frac{1}{2}Diag(1,-1,0,0)$, $D_{2}=T_{8}=\frac{1}{2\sqrt{3}}Diag(1,1,-2,0)$,
and $D_{3}=T_{15}=\frac{1}{2\sqrt{6}}Diag(1,1,1,-3)$ respectively.
In this basis, the gauge fields are $A_{\mu}^{0}$ and $A_{\mu}\in su(4)$,
that is \begin{equation}
A_{\mu}=\frac{1}{2}\left(\begin{array}{cccc}
D_{\mu}^{1} & \sqrt{2}X_{\mu} & \sqrt{2}X_{\mu}^{\prime} & \sqrt{2}K_{\mu}\\
\\\sqrt{2}X_{\mu}^{*} & D_{\mu}^{2} & \sqrt{2}W_{\mu} & \sqrt{2}K_{\mu}^{\prime}\\
\\\sqrt{2}X_{\mu}^{\prime*}{} & \sqrt{2}W_{\mu}^{*} & D_{\mu}^{3} & \sqrt{2}Y_{\mu}\\
\\\sqrt{2}K_{\mu}^{*} & \sqrt{2}K_{\mu}^{\prime*} & \sqrt{2}Y_{\mu}^{*} & D_{\mu}^{4}\end{array}\right),\label{Eq.7}\end{equation}
 with neutral diagonal bosons: $D_{\mu}^{1}=A_{\mu}^{3}+A_{\mu}^{8}/\sqrt{3}+A_{\mu}^{15}/\sqrt{6}$,
$D_{\mu}^{2}=-A_{\mu}^{3}+A_{\mu}^{8}/\sqrt{3}+A_{\mu}^{15}/\sqrt{6}$,
$D_{\mu}^{3}=-2A_{\mu}^{8}/\sqrt{3}+A_{\mu}^{15}/\sqrt{6}$, and $D_{\mu}^{4}=-3A_{\mu}^{15}/\sqrt{6}$
respectively. Apart from the charged Weinberg bosons, $W$, there
are new charged bosons, $K$, $K^{\prime}$, $X$, $X^{\prime}$ and
$Y$. Note that $X$ is doubly charged coupling different chiral states
of the same charged lepton (the so called ''bilepton''), while $Y$
is neutral.

The next step is the diagonalization of the mass matrix Eq. (33) in
Ref. \cite{key-24}. By that parameter choice - namely, matrix $\eta$
in the scalar sector - and by imposing that $m(Z)=m(W)/\cos\theta_{W}$
(from the SM) is eigenvalue of that marix, one gets the matrix $\omega$.
It is:

\begin{equation}
\omega=\left(\begin{array}{ccccc}
\frac{\sqrt{1-4\sin^{2}\theta_{W}}}{2\cos\theta_{W}} &  & -\frac{\sqrt{3}}{2\cos\theta_{W}} &  & 0\\
\\\frac{\sqrt{3}}{2\cos\theta_{W}} &  & \frac{\sqrt{1-4\sin^{2}\theta_{W}}}{2\cos\theta_{W}} &  & 0\\
\\0 &  & 0 &  & 1\end{array}\right).\label{Eq.8}\end{equation}

\paragraph{Neutral Charges}

Evidently, the two subject to diagonalization are - $Z^{1}=Z$ of
the SM, and $Z^{2}=Z^{\prime}$ respectively. They exhibit the following
neutral charges (after working out in Eq.(\ref{Eq.2}) the versor
choice $\nu_{1}=1$, $\nu_{2}=0$, $\nu_{3}=0$ and the coupling matching
$g^{\prime}/g=\sin\theta_{W}/\sqrt{1-4\sin^{2}\theta_{W}}$ and $\sin\theta=2\sin\theta_{W}$):

\begin{equation}
Q^{\rho}(Z^{1})=g\left[\left(D_{1}^{\rho}\sqrt{1-4\sin^{2}\theta_{W}}-y_{ch}^{\rho}\frac{2\sin^{2}\theta_{W}}{\sqrt{1-4\sin^{2}\theta_{W}}}\right)\omega_{\cdot\;1}^{1\;\cdot}+D_{2}^{\rho}\omega_{\cdot\;1}^{2\;\cdot}\right],\label{Eq.9}\end{equation}

\begin{equation}
Q^{\rho}(Z^{2})=g\left[\left(D_{1}^{\rho}\sqrt{1-4\sin^{2}\theta_{W}}-y_{ch}^{\rho}\frac{2\sin^{2}\theta_{W}}{\sqrt{1-4\sin^{2}\theta_{W}}}\right)\omega_{\cdot\;2}^{1\;\cdot}+D_{2}^{\rho}\omega_{\cdot\;2}^{2\;\cdot}\right],\label{Eq.10}\end{equation}
 while the heaviest neutral boson - $Z^{3}=Z^{\prime\prime}$, in
our notation - will couple the fermion representations through:

\begin{equation}
Q^{\rho}(Z^{3})=gD_{3}^{\rho}.\label{Eq.11}\end{equation}

Now, it is a matter of some algebra - assuming that $\omega_{\cdot\;1}^{1\;\cdot}=\omega_{\cdot\;2}^{2\;\cdot}=\frac{\sqrt{1-4\sin^{2}\theta_{W}}}{2\cos\theta_{W}}$
and $\omega_{\cdot\;1}^{2\;\cdot}=-\omega_{\cdot\;2}^{1\;\cdot}=-\frac{\sqrt{3}}{2\cos\theta_{W}}$
along with the particular assignments of the fermion representations
($\rho$) - Eqs. (\ref{Eq.3}) - (\ref{Eq.6}) - and the neutral charges
of all fermions in the 3-4-1 model of interest here are computed.
They are sumarized in the Table. We took into account that at the
SM level $e=g\sin\theta_{W}$ holds.

\begin{table}

\caption{Coupling coefficients of the neutral currents in 3-4-1 model with
exotic electric charges}

\begin{tabular}{cccccc}
\hline 
Particle\textbackslash{}Coupling($e/\sin2\theta_{W}$)&
$Z\rightarrow\bar{f}f$&
&
$Z^{\prime}\rightarrow\bar{f}f$&
&
$Z^{\prime\prime}\rightarrow\bar{f}f$\tabularnewline
\hline
\hline 
&
&
&
&
&
\tabularnewline
&
&
&
&
&
\tabularnewline
$\nu_{eL},\nu_{\mu L},\nu_{\tau L}$&
$1$&
&
$-\frac{\sqrt{1-4\sin^{2}\theta_{W}}}{\sqrt{3}}$&
&
$\frac{\cos\theta_{W}}{\sqrt{6}}$\tabularnewline
&
&
&
&
&
\tabularnewline
$e_{L},\mu_{L},\tau_{L}$&
$2\sin^{2}\theta_{W}-1$&
&
$-\frac{\sqrt{1-4\sin^{2}\theta_{W}}}{\sqrt{3}}$&
&
$\frac{\cos\theta_{W}}{\sqrt{6}}$\tabularnewline
&
&
&
&
&
\tabularnewline
$e_{R},\mu_{R},\tau_{R}$&
$-2\sin^{2}\theta_{W}$&
&
$\frac{2\sqrt{1-4\sin^{2}\theta_{W}}}{\sqrt{3}}$&
&
$\frac{\cos\theta_{W}}{\sqrt{6}}$\tabularnewline
&
&
&
&
&
\tabularnewline
$N_{eL},N_{\mu L},N_{\tau L}$&
$0$&
&
$0$&
&
$-\sqrt{\frac{3}{2}}\cos\theta_{W}$\tabularnewline
&
&
&
&
&
\tabularnewline
$u_{L},c_{L}$&
$1-\frac{4}{3}\sin^{2}\theta_{W}$&
&
$\frac{1}{\sqrt{3}}\left(\frac{1-2\sin^{2}\theta_{W}}{\sqrt{1-4\sin^{2}\theta_{W}}}\right)$&
&
$-\frac{\cos\theta_{W}}{\sqrt{6}}$\tabularnewline
&
&
&
&
&
\tabularnewline
$d_{L},s_{L}$&
$-1+\frac{2}{3}\sin^{2}\theta_{W}$&
&
$\frac{1}{\sqrt{3}}\left(\frac{1-2\sin^{2}\theta_{W}}{\sqrt{1-4\sin^{2}\theta_{W}}}\right)$&
&
$-\frac{\cos\theta_{W}}{\sqrt{6}}$\tabularnewline
&
&
&
&
&
\tabularnewline
$t_{L}$&
$1-\frac{4}{3}\sin^{2}\theta_{W}$&
&
-$\left(\frac{1}{\sqrt{3}}\right)\frac{1}{\sqrt{1-4\sin^{2}\theta_{W}}}$&
&
$\frac{\cos\theta_{W}}{\sqrt{6}}$\tabularnewline
&
&
&
&
&
\tabularnewline
$b_{L}$&
$-1+\frac{2}{3}\sin^{2}\theta_{W}$&
&
$-\left(\frac{1}{\sqrt{3}}\right)\frac{1}{\sqrt{1-4\sin^{2}\theta_{W}}}$&
&
$\frac{\cos\theta_{W}}{\sqrt{6}}$\tabularnewline
&
&
&
&
&
\tabularnewline
$T_{L}$&
$-\frac{4}{3}\sin^{2}\theta_{W}$&
&
$-\left(\frac{4}{\sqrt{3}}\right)\frac{\sin^{2}\theta_{W}}{\sqrt{1-4\sin^{2}\theta_{W}}}$&
&
$-\sqrt{\frac{3}{2}}\cos\theta_{W}$\tabularnewline
&
&
&
&
&
\tabularnewline
$D_{1L},D_{2L}$&
$\frac{2}{3}\sin^{2}\theta_{W}$&
&
$\left(\frac{2}{\sqrt{3}}\right)\frac{\sin^{2}\theta_{W}}{\sqrt{1-4\sin^{2}\theta_{W}}}$&
&
$\sqrt{\frac{3}{2}}\cos\theta_{W}$\tabularnewline
&
&
&
&
&
\tabularnewline
$u_{R},c_{R},t_{R},T{}_{R}$&
$-\frac{4}{3}\sin^{2}\theta_{W}$&
&
$\left(\frac{4}{\sqrt{3}}\right)\frac{\sin^{2}\theta_{W}}{\sqrt{1-4\sin^{2}\theta_{W}}}$&
&
$0$\tabularnewline
&
&
&
&
&
\tabularnewline
$d_{R},s_{R},b_{R},D_{iR}$&
$+\frac{2}{3}\sin^{2}\theta_{W}$&
&
$-\left(\frac{2}{\sqrt{3}}\right)\frac{\sin^{2}\theta_{W}}{\sqrt{1-4\sin^{2}\theta_{W}}}$&
&
$0$\tabularnewline
&
&
&
&
&
\tabularnewline
$J_{1L},J_{2L}$&
$\frac{8}{3}\sin^{2}\theta_{W}$&
&
$-\left(\frac{2}{\sqrt{3}}\right)\frac{1-5\sin^{2}\theta_{W}}{\sqrt{1-4\sin^{2}\theta_{W}}}$&
&
-$\frac{\cos\theta_{W}}{\sqrt{6}}$\tabularnewline
&
&
&
&
&
\tabularnewline
$J_{1R},J_{2R}$&
$\frac{8}{3}\sin^{2}\theta_{W}$&
&
$-\left(\frac{8}{\sqrt{3}}\right)\frac{\sin^{2}\theta_{W}}{\sqrt{1-4\sin^{2}\theta_{W}}}$&
&
$0$\tabularnewline
&
&
&
&
&
\tabularnewline
$J_{3L}$&
$-\frac{10}{3}\sin^{2}\theta_{W}$&
&
$\left(\frac{2}{\sqrt{3}}\right)\frac{1-6\sin^{2}\theta_{W}}{\sqrt{1-4\sin^{2}\theta_{W}}}$&
&
$\frac{\cos\theta_{W}}{\sqrt{6}}$\tabularnewline
&
&
&
&
&
\tabularnewline
$J_{3R}$&
$-\frac{10}{3}\sin^{2}\theta_{W}$&
&
$\left(\frac{10}{\sqrt{3}}\right)\frac{\sin^{2}\theta_{W}}{\sqrt{1-4\sin^{2}\theta_{W}}}$&
&
$0$\tabularnewline
&
&
&
&
&
\tabularnewline
&
&
&
&
&
\tabularnewline
\hline 
&
&
&
&
&
\tabularnewline
\end{tabular}
\end{table}

\section{Conclusions}

In this letter, we have presented in detail the neutral currents of
a particular 3-4-1 gauge model based on the general approach proposed
by Cot\u{a}escu \cite{key-23}. As one can observe by inspecting
the Table, all the SM values are naturally recovered, while the two
new neutral bosons of this model ehibit particular features. Namely,
$Z^{\prime}$is leptophobic, since $\sin\theta_{W}\simeq0.223$, but
supplies very large couplings for quarks. The third neutral boson
$Z^{\prime\prime}$ seems to make no distinction among the SM fermions,
being at the same time quite a vector like interaction with respect
to the SM fermions, while for the exotic particles it is strongly
parity-violating. 

These results are the exact couplings of the fermion and they do not
need any supplemental mixing angle since this job have been done as
an inner step by the method itself through the general Weinberg transformation.
The resulting values are important for some specific decays and weak
processes, but these computations are beyond the scope of this letter.

\end{document}